\documentclass[aps,showpacs,preprintnumbers,amsmath,amssymb,pra,twocolumn,superscriptaddress
]{revtex4-1}
\usepackage{txfonts}
\usepackage{float}
\usepackage{graphicx}
\usepackage{dcolumn}
\usepackage{bm}
\usepackage{color}
\usepackage{subfigure}  
\def\comment#1{}

\def\slashchar#1{\setbox0=\hbox{$#1$}           
	\dimen0=\wd0                                 
	\setbox1=\hbox{/} \dimen1=\wd1               
	\ifdim\dimen0>\dimen1                        
	\rlap{\hbox to \dimen0{\hfil/\hfil}}      
	#1                                        
	\else                                        
	\rlap{\hbox to \dimen1{\hfil$#1$\hfil}}   
	/                                         
	\fi}                                         %

\def\sigmab{{\mbox{\boldmath $\sigma$}}}
\def\nablab{{\mbox{\boldmath $\nabla$}}}

\def\Phib{{\mbox{\boldmath $\Phi$}}}

\begin{document}

\title{First and second-order metal-insulator phase transitions and topological aspects of a Hubbard-Rashba system}

\author{Edgar Marcelino}
\affiliation{
Institut f\"ur Theoretische Physik III, Ruhr-Universit\"at Bochum,
Universit\"atsstrasse 150, DE-44801 Bochum, Germany}

\begin{abstract}
This paper considers a model consisting of a kinetic term, Rashba spin-orbit coupling and short-range Coulomb interaction at zero-temperature. The Coulomb interaction is decoupled by a mean-field approximation in the spin channel using field theory methods. The results feature a first-order phase transition for any finite value of the chemical potential and quantum criticality for vanishing chemical potential. The Hall conductivity is also computed using Kubo formula in a mean-field effective Hamiltonian. In the limit of infinite mass the kinetic term vanishes and all the phase transitions are of second order, in this case spontaneous symmetry breaking mechanism adds a ferromagnetic metallic phase to the system and features a zero-temperature quantization of the Hall conductivity in the insulating one.
\end{abstract}

\maketitle

\section{Introduction}

Two-dimensional spin-orbit coupled systems are typically well described by spin-orbit interactions (SOIs) of Rashba and Dresselhaus types 
\cite{Dresselhaus,Rashba,SOI}, which play an important role  in many aspects of semiconductor physics,   
such as spin interferometer \cite{Interf1,Interf2}, persistent spin helix \cite{Helix1,Helix2}, and classical 
and quantum spin Hall effects \cite{Hall1,Hall2, Hall3, Kane_TI2D,Kane,Zhang_TI2D}. SOI is closely related to interesting properties of materials yielding different kinds of topological orders, this ensures the presence of robust metallic edge states due to the impossibility for the system to interpolate between two phases with different topological orders without closing the gap. 
The first example of this kind of topological order appearing in Condensed Matter Physics was given by the integer quantum Hall effect \cite{Klitizing}, characterized by the TKNN index \cite{TKNN} and featuring a quantization of the Hall conductivity related to the Chern-numbers of the bands located below the chemical potential. The ideas in \cite{TKNN}, first restricted to lattice systems, were later generalized to continuous systems and obtained in situations where a strong transverse magnetic field, such as in \cite{Klitizing}, could be absent. Indeed, one main ingredient for the manifestation of this topological order in two-dimensional systems is the breaking of time reversal (TR) symmetry \cite{Haldane} and those analogous phases of matter, that do not require any 
external magnetic field at all, are the so called Chern insulators or anomalous quantum Hall effect \cite{Chern-Insulators}. 
When TR symmetry is present another kind of topological order, characterized by a $Z_{2}$ topological index, may manifest. This was first proposed in \cite{Kane_TI2D,Kane}, yielding a so called topological insulator (TI). In this case spin is locked to momentum leading to TR symmetry-protected gapless edge states coexisting with gapped bulk states \cite{Hasan-Kane-RMP,Zhang-RMP-2011}, which is provided by the band structure of several materials, even in absence of electron-electron interactions \cite{Zhang-RMP-2011}. When TR is present in two-dimensional systems, charge-carriers with opposite spin flow in opposite directions,  
featuring a quantized spin Hall conductivity, with a vanishing Hall conductivity being ensured after considering the contribution from both spins.

One important example of interacting system including SOI that exhibits a rich phase diagram is the so called Kane-Mele-Hubbard model \cite{Stephan-Rachel-1}, where an SOI of the Rashba type is implemented in a Honeycomb lattice. In this case the phase diagram  features a quantum spin-Hall phase and also a Mott insulating antiferromagnetic one \cite{Stephan-Rachel-1,Stephan-Rachel-2,Assaad-KMH}. There was also a discussion about a spin liquid phase in the Kane-Mele Hubbard model \cite{Assaad-KMH}, but this idea was rejected later \cite{Sorella-KMH,Assaad-KMH-2}.  Such a rich phase structure also arises in the square lattice for the Bernevig-Hughes-Zhang (BHZ) model \cite{BHZ} when Hubbard interactions are included \cite{Tada,Yoshida,Miyakoshi,Budich}.
A considerably simpler framework allowing an explicit analytic treatment is a Rashba-Hubbard system defined directly as a continuum field theory, with a second-quantized Hamiltonian given by ($\hbar=v_{F}=1$),
\begin{equation}
\label{Eq:Rashba-Hubbard}
{\cal H}=\frac{1}{2m}\nablab\psi^\dagger\cdot\nablab\psi-\mu\psi^\dagger\psi-i\alpha\psi^\dagger(\nablab\times\hat {z})\cdot\sigmab \psi
-\frac{2g}{3}{\bf S}^2,
\end{equation} 
 where $\psi^\dagger=[\psi_\uparrow^\dagger~\psi^\dagger_\downarrow]$, $\sigmab=(\sigma_x,\sigma_y,\sigma_z)$ is a vector of 
 Pauli matrices, and ${\bf S}=(1/2)\psi^\dagger\sigmab\psi$.  The last term is just a repulsive Hubbard interaction in disguise, since 
 ${\bf S}^2=(3/2)\psi^\dagger \psi-(3/4)(\psi^\dagger\psi)^2$. 

In this paper this model is studied using a mean-field approximation within an effective action formalism obtained by means of a Hubbard-Stratonovich transformation in the spin channel and integrating out the fermionic fields. The mean-field results feature a first-order metal-insulator phase transition for any finite value of the chemical potential. For a vanishing 
chemical potential a second-order phase transition is obtained.  The Hall conductivity of the system is also computed and it remains finite only in the insulating phase. Since the system is gapless in the metallic phase, no quantization of the Hall conductivity 
is obtained in this case.  The limit of infinite mass, in which case the kinetic term may be neglected, is also studied and leads to a model that is TR invariant only in the paramagnetic phase. In this case spontaneous TR symmetry breaking leads to a metallic ferromagnetic phase in addition to the paramagnetic and ferromagnetic insulating ones. The Hall conductivity varies linearly with the gap in the metallic phase, being quantized in the insulating regime, where it assumes only two distinct values. In the paramagnetic case the system has TR symmetry and the Hall conductivity vanishes, as expected.

The results obtained here are quite distinct from other ones following from standard models of magnetism in the literature. The Hubbard model 
typically yields a second-order phase transition, but not a first-order one. The Stoner model does not even exhibit a phase transition at zero temperature. Despite the fact that the results in this paper are quite different from others in literature, they resemble partially the ones from a discontinuous metal-insulator transition proposed by Mott in which he considers an array of hydrogen-like atoms with arbitrary lattice constant. In Mott's paper it is assumed that the transition between two states, with all electrons being trapped or free, would occur when a screened Yukawa-like potential round each constant. The screening behavior is determined by the Thomas-Fermi method. Details about the Mott transition can be founded in Ref. \cite{Mott} and references therein. Mott also studied later whether, or not, this phase transition would become continuous in the presence of disorder. 

In the next section the Lagrangian density of the system with a chemical potential is considered and an effective action at finite temperature is obtained after performing a Hubbard-Stratonovich transformation. The saddle-point equations for the gap and charge density are derived after integrating 
out the fermions. 
In section III the mean-field equations are solved and the phase transitions of the system are studied at zero  temperature and also in the limit of infinite mass. In section IV the Hall conductivity is computed by considering a mean-field effective Hamiltonian and using the Kubo formula. The last section contains the conclusions and some discussions about future perspectives for this work.

\section{Effective action and gap equation} 

In order to study the model at the mean-field level and obtain systematic fluctuation corrections, it is more convenient to use a 
functional integral formalism where the fermionic operators are replaced by Grassmann fields in a standard fashion. In this case, upon 
performing a Hubbard-Stratonovich (HS) transformation, a formally quadratic action in the fermionic fields is obtained. Using an imaginary 
time formalism at finite temperature, the Lagrangian featuring the HS field, $\Phib$, is obtained as, 
\begin{equation}
\label{Eq:L}
{\cal L}=\psi^\dagger\left[\partial_\tau-\mu+H(\Phib)\right]\psi+\frac{1}{2}\Phib^2,
\end{equation}
where $H$ represents the Hamiltonian operator,
\begin{equation}
\label{Eq:H}
H(\Phib)=-\frac{\nabla^2}{2m}-\left(i\alpha\nablab\times\hat{\bf z}+
\sqrt{\frac{g}{3}}\Phib\right)\cdot\sigmab.
\end{equation}
After performing the Gaussian integral in the Grassmann fields, the {\it exact} effective action is given by, 
\begin{equation} \label{Seff}
S_{\rm eff}(\Phib)=-{\rm Tr}\ln\left[\partial_\tau-\mu+H(\Phib)\right]+\frac{1}{2}\int_0^\beta d\tau\int d^2r~\Phib^2.
\end{equation}
A mean-field theory is obtained by assuming $\Phib$ to be uniform within a lowest order approximation, in which case 
the tracelog above can be evaluated more easily.  Indeed, in this case, 
\begin{equation}  \label{Tracelog}
\frac{1}{A}{\rm Tr}\ln\left[\partial_\tau-\mu+H(\Phib)\right]\approx\beta\sum_{\sigma=\pm}\int\frac{d^2k}{(2\pi)^2}
\ln\left[1+e^{-\beta E_\sigma({\bf k})}\right],
\end{equation}
where $A$ is an (infinite) area, and,  
\begin{equation} \label{Energy_Spec}
E_{\pm}(\mathbf{k})=\frac{\mathbf{k}^{2}}{2m}-\mu \pm \eta(\mathbf{k}),
\end{equation}
with 
\begin{equation}
\eta(\mathbf{k})=\sqrt{\left( \alpha k_{x}-\sqrt{\frac{g}{3}} \Phi_{y} \right)^{2}+\left( \alpha k_{y}+\sqrt{\frac{g}{3}} \Phi_{x} \right)^{2}+\frac{g}{3}\Phi_{z}^{2}}.
\end{equation}

The mean-field theory is determined by the "equation of motion" for $\Phib$, i.e., 
\begin{equation}
\Phi_j=2\sqrt{\frac{g}{3}}\langle S_j\rangle.
\end{equation}
Thus, using the above result in the Heisenberg equations of motion for the electronic spin, the following equations are derived, 
\begin{equation}
\partial_t\langle S_x\rangle=2\alpha k_x\langle S_x\rangle,
\end{equation}
\begin{equation}
\partial_t\langle S_y\rangle=2\alpha k_y\langle S_y\rangle,
\end{equation}
\begin{equation}
\partial_t\langle S_z\rangle=2\alpha {\bf k}\cdot\langle {\bf S}\rangle.
\end{equation}
Therefore, in order to have $\Phib$ also uniform in time, we need $\Phi_x=\Phi_y=0$, such that only $\Phi_z=\phi$ is nonzero.  Under this assumption, the free energy density is minimized to obtain the gap equation,
\begin{equation} \label{gap_equation}
\phi=\frac{g \phi}{6 \pi} \int_{0}^{\infty} \frac{f_{-}(k)-f_{+}(k)}{\epsilon(k)}kdk, 
\end{equation}
where $f_{\pm}(k)=1/[e^{\beta (k^{2}/2m-\mu \pm \epsilon(k))}+1]$ and $\epsilon(k)=\sqrt{\alpha^{2} k^{2}+(g/3)\phi^{2}}$. The charge density can also be easily computed from the free energy and leads to,
\begin{equation} \label{density_equation}
n=\frac{1}{2 \pi}  \int_{0}^{\infty} \left(f_{-}(k)+f_{+}(k) \right)kdk, 
\end{equation}

In the next section the zero temperature solutions of the saddle-point equations (Eq. \ref{gap_equation} and Eq. \ref{density_equation}) will be analytically derived and the corresponding phase transitions will be discussed.

\section{First-order phase transition and quantum critical point}

At zero temperature a nonzero value of the gap, corresponding to a ferromagnetic solution, satisfies, 
\begin{eqnarray} 
\label{mean-field(T=0)}
\frac{6 \pi}{g}=\sum_{\sigma= \pm 1} \sigma \int_{0}^{\infty} \frac{kdk}{\epsilon(k)} \Theta \left(\mu-\frac{k^{2}}{2m}+\sigma \epsilon(k) \right),
\end{eqnarray}
where $\Theta(x)$ is the Heaviside function. The charge density can also be evaluated in the same way,
\begin{eqnarray} 
\label{mean-field(T=0)2}
n &=& \sum_{\sigma= \pm 1} \int_{0}^{\infty} \frac{kdk}{2 \pi} \Theta \left(\mu-\frac{k^{2}}{2m}+\sigma \epsilon(k) \right).
\end{eqnarray}

Denoting $M=\phi \sqrt{\frac{g}{3}}$ and performing 
the integrals above, one may obtain for the gap equation,    
\begin{equation} 
\label{gap_T=0}
\frac{6 \pi\alpha^2}{g}=
\begin{cases}
m\alpha^2-|M|+\sqrt{m^2\alpha^4+2m\alpha^2\mu+|M|^2} & (\mu \leq |M|) \\
2m\alpha^2 & (\mu \geq |M|)
\end{cases}
\end{equation}
and for the charge density,
\begin{equation} 
\label{density_T=0}
\frac{2 \pi n}{m} =
\begin{cases}
m\alpha^2+\mu+\sqrt{m^2\alpha^4+2m\alpha^2\mu+|M|^2} & (\mu \leq |M|) \\
2(\mu+m\alpha^2) & (\mu \geq |M|).
\end{cases}
\end{equation}

Using Eq. (\ref{density_T=0}) one can write the chemical potential as a function of the charge density in the insulating phase $(\mu \leq |M|)$ and in the metallic one $(\mu \geq |M|)$. At first it may seem that there will be more than one solution in each case but since the chemical potential and the density should be positive and respect the corresponding inequality of each phase, given by the Heaviside functions, it is possible to find a unique solution and comparison with Eq. (\ref{gap_T=0}) leads to,
\begin{equation} 
\label{g_T=0}
\frac{6 \pi\alpha^2}{g}=\begin{cases}2m \alpha^{2} &  \left( \frac{\pi n}{m}-m \alpha^{2}-|M| \geq 0 \right) \\
\sqrt{4 \pi n \alpha^{2}+M^{2}}-|M| & \left(|M|-\frac{\pi n}{m} +m \alpha^{2} \geq 0 \right),
\end{cases}
\end{equation}
where $\mu=\frac{\pi n}{m}-m \alpha^{2} \geq |M| \geq 0$ in the metallic phase and $|M|-\frac{\pi n}{m} +m \alpha^{2} \geq 0$ in the insulating one $\left(\mu= \frac{2 \pi n}{m}-\sqrt{4 \pi n \alpha^{2}+M^{2}} \leq |M| \right)$. 

A quick look to Eq. (\ref{g_T=0}) shows an apparent independence of the gap with respect to the coupling constant in the metallic phase, which doesn't seem to be reasonable. To understand that better one can write $\phi$ as a function of $g$ in the insulator phase and notice that $g=g_c=3 \pi/m$ only for $\phi=M=0$, which corresponds to the value of the coupling constant in the metallic phase. The conclusion is that $\phi=0$ for $g<g_c$ and the metallic phase coincides with the paramagnetic one. Thus, Eq. (\ref{g_T=0}) can be inverted,
\begin{equation} 
\label{M_T=0}
|M|=\left( \frac{gn}{3}-\frac{3 \pi \alpha^{2}}{g} \right) \Theta \left(g-g_{c} \right).
\end{equation}

\begin{figure}[h!]
	\centering    
	\includegraphics[width=0.5\textwidth]{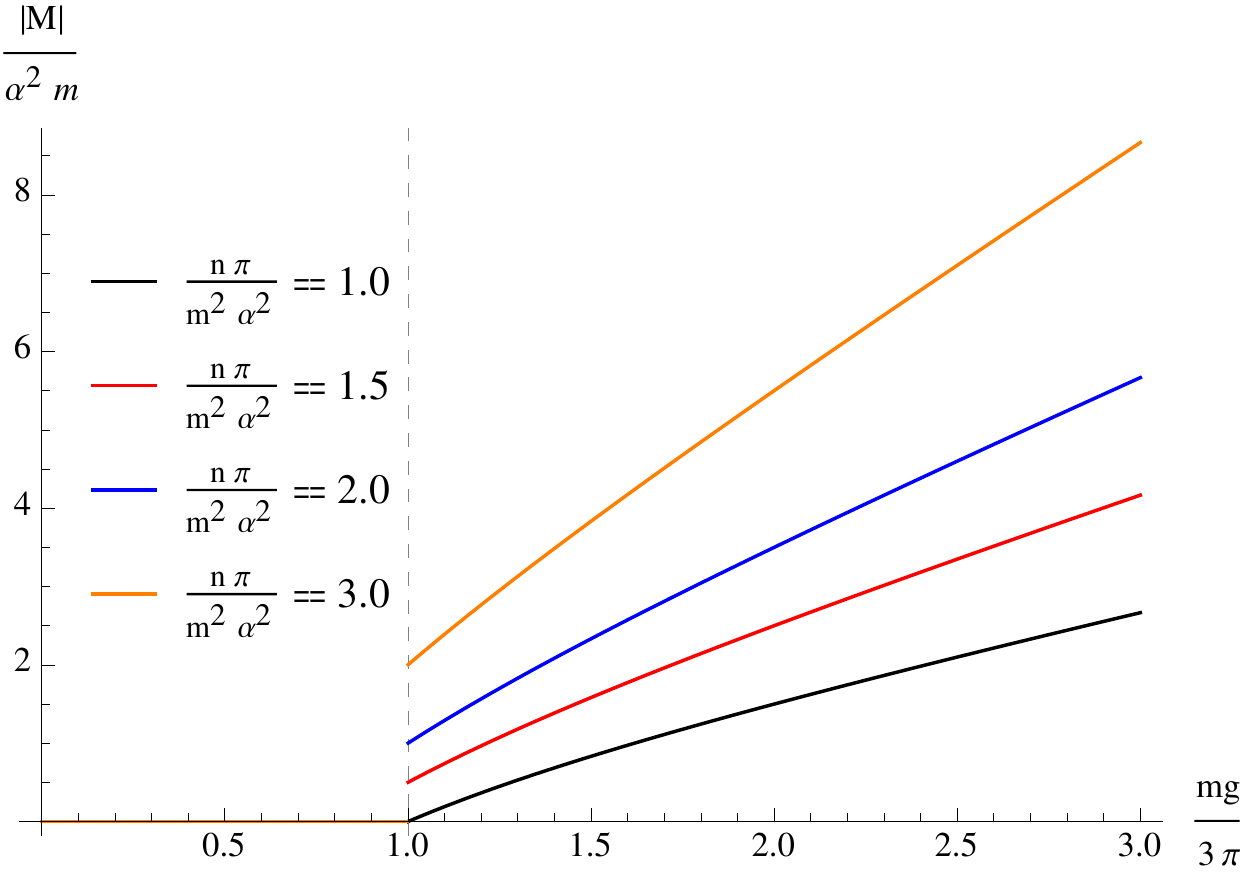}
	\caption{(Color online)  Behavior of the gap [Eq. (\ref{M_T=0})] with the coupling constant $g$ for different values of the density. The black line corresponding to $\alpha=\alpha_c=\frac{\pi n}{m^{2}}$ leads to $\mu=0$ and shows a second order phase transition where $g=g_c=\frac{3 \pi}{m}$ is a quantum critical point.}
	\label{phi_x_g(T=0)}
\end{figure}

The solution on Eq. (\ref{M_T=0}) is interesting because it yields for $\mu>0$ a first-order metal-insulator transition where the insulating phase is ferromagnetic. For $\mu=0$ it features a second-order phase transition and $g=g_c$ becomes a quantum critical point. The behavior of the gap with the coupling constant $g$ is shown in Fig. \ref{phi_x_g(T=0)}. 

\begin{figure}[h!]
	\centering    
	\includegraphics[width=0.5\textwidth]{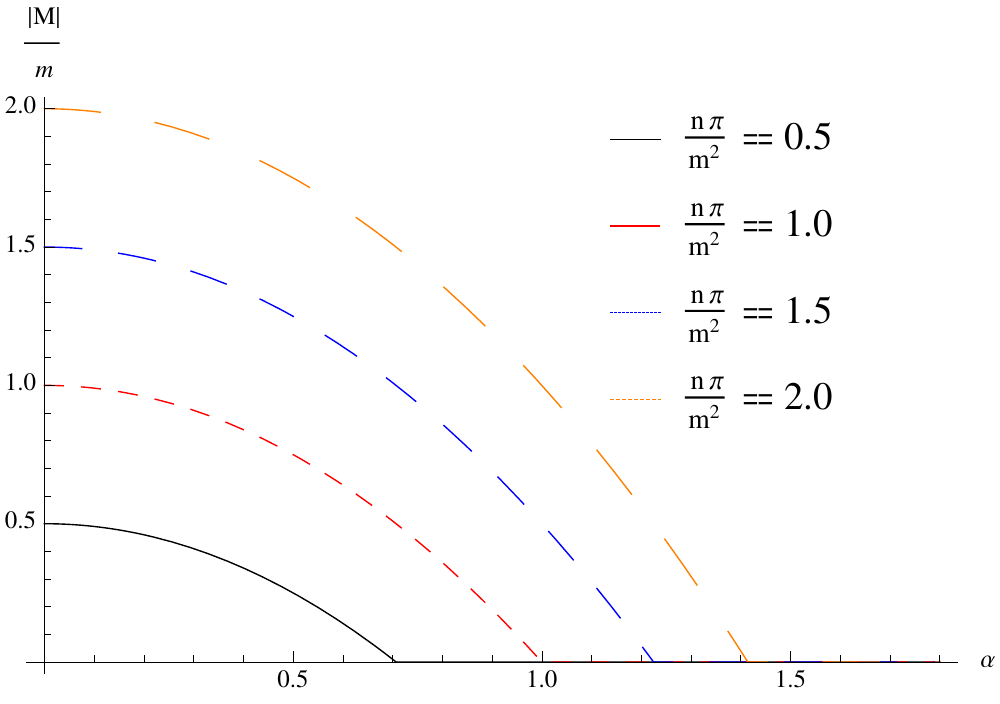}
	\caption{(Color online)  Behavior of the gap [Eq. (\ref{M_T=0})] with the spin-orbit coupling constant $\alpha$ for different values of the density and $\frac{mg}{3 \pi}=1.0$.}
	\label{M_x_a_T=0}
\end{figure}

For $g = g_c$ the system is in the gapless case and with null chemical potential $(\mu=M=0)$, thus the critical spin-orbit coupling is completely determined as it is easy to see from Eq. (\ref{density_T=0}) or from Eq. (\ref{M_T=0}) and leads to $ \alpha^2=\alpha_{c}^{2}=\frac{\pi n}{m^{2}}$. For $g  \geq g_{c}$ the magnetization decreases with $\alpha$ until vanishing as showed in Fig. \ref{M_x_a_T=0} for $\frac{mg}{3 \pi}=1.0$, notice that the magnetization reaches zero exactly at $\alpha=\alpha_{c}=\frac{\pi n}{m^{2}}$.

%
%
A limit case of interest is the one where the SOI is very strong, being formally given by the infinite mass limit ($m\to\infty$). In this regime the system is TR invariant only in the paramagnetic phase $(\phi=0)$ and spontaneous symmetry breaking allows an additional ferromagnetic metallic phase. Furthermore, as it will now be shown, this case exhibits a behavior quite distinct from the one given by Eq. (\ref{gap_T=0}), as it always features a quantum critical point. The reason for these differences lies on the fact that the limit $m\to\infty$ is singular. Indeed, 
in order to evaluate the integrals in Eq. (\ref{mean-field(T=0)}) and Eq. (\ref{mean-field(T=0)2}) for $m\to\infty$ we have to introduce a large momentum cutoff ($\Lambda$),
\begin{eqnarray} \label{gap_iso(T=0)}
\frac{6 \pi\alpha^{2}}{g}=\int_{|M|}^{\sqrt{\alpha^{2} \Lambda^{2}+M^{2}}} [\Theta(\mu+\epsilon)-\Theta(\mu-\epsilon)]d \epsilon,  \\  \label{n_iso(T=0)}
2 \pi n \alpha^{2}= \int_{|M|}^{\sqrt{\alpha^{2} \Lambda^{2}+M^{2}}} [\Theta(\mu+\epsilon)+\Theta(\mu-\epsilon)] \epsilon d \epsilon.
\end{eqnarray}
The metallic and insulating regimes will be analyzed separately, since they describe distinct critical behavior. The insulating regime will be studied at first.  
Assuming $|M|\ll \alpha\Lambda$, Eq. (\ref{gap_iso(T=0)}) and Eq. (\ref{n_iso(T=0)}) reduce to, 
\begin{eqnarray}
\frac{6\pi\alpha^2}{g}=\alpha\Lambda-|M|, \\
n=\frac{\Lambda^{2}}{4 \pi},
\end{eqnarray}
which yields, 
\begin{eqnarray}
|M|=6\pi\alpha^2\left(\frac{1}{g_{ci}}-\frac{1}{g}\right), \\
n=\frac{9 \pi \alpha^{2}}{g_{ci}^{2}}, \label{n_TI1}
\end{eqnarray}
where $g_{ci}=6\pi\alpha/\Lambda$ is the critical point for the insulating regime. Since $|M|\geq 0$, we have that the gap vanishes for 
$g\leq g_{ci}$. 
  
On the other hand, for the ordered metallic regime Eq. (\ref{gap_iso(T=0)}) and Eq. (\ref{n_iso(T=0)}) become,    
\begin{eqnarray}
\frac{6\pi\alpha^2}{g}=\sqrt{\alpha^{2} \Lambda^{2}+M^{2}}-\mu, \\
n=\frac{1}{4 \pi \alpha^{2}} \left[\alpha^{2} \Lambda^{2} + \mu^{2}-M^{2}\right],
\end{eqnarray}
and the quantum critical point is now located at, 
\begin{equation}
\label{Eq:gc-m_large}
g_{cm}=\frac{6\pi\alpha^2}{\alpha\Lambda-\mu}, 
\end{equation}
such that for $g\geq g_{cm}$, 
\begin{eqnarray}
|M|=\sqrt{\left(\frac{6\pi\alpha^2}{g}+\mu \right)^{2}-\left(\frac{6\pi\alpha^2}{g_{cm}}+\mu \right)^{2}}, \\
n=\frac{1}{4 \pi \alpha^{2}} \left[ \left( \frac{6\pi\alpha^2}{g_{cm}}+\mu \right)^{2}+\mu^{2}-M^{2} \right]. \label{n_TI2}
\end{eqnarray}
Note that for $M=0$ and $\mu \rightarrow 0^{+}$, the critical couplings in each regime become the same ($g_{cm}=g_{ci}$) and Eq. (\ref{n_TI2}) reduces to Eq. (\ref{n_TI1}). The chemical potential does not need to be small as compared to $\alpha\Lambda$ in the metallic regime, but Eq. (\ref{Eq:gc-m_large}) requires 
$\mu<\alpha\Lambda$, in order to keep $g_{cm}$ positive.  
Thus, the ferromagnetic order parameter in the metallic case exhibits a different power law from the ferromagnetic
insulating one.   This jump in the critical behavior  of the order parameter replaces the jump in the order parameter found in the 
finite $m$ case, reflecting the Fermi surface singularity of the metallic phase.

\section{Topological charge and Hall conductivity}

From Eq. (\ref{Eq:L}) and Eq. (\ref{Eq:H}) it is possible to write a mean-field effective Hamiltonian in the form, 
\begin{equation} \label{H_mean-field}
H=\epsilon(k) I+\mathbf{d}\cdot \sigmab
\end{equation}
where $\epsilon(k)=\frac{\mathbf{k}^{2}}{2m}-\mu+\frac{\phi^{2}}{2}$ and $\mathbf{d}=(\alpha k_{y},- \alpha k_{x},M)$, leading to the following energy spectrum:
\begin{eqnarray} \label{Energy_Spectrum}
E_{\pm}(\mathbf{k})=\frac{\mathbf{k}^{2}}{2m}-\mu \pm \sqrt{\alpha^{2} \mathbf{k}^{2}+M^{2}}.
\end{eqnarray}
For a Hamiltonian such as the one given by Eq. (\ref{H_mean-field}) the 
Kubo formula yields a Hall conductivity $\sigma_{xy}$ \cite{Qi(Kubo_formula)}, 
\begin{equation} \label{Hall}
\sigma_{xy}(T,M)=\frac{e^{2}}{2h} \int d^{2} \mathbf{k} Q_{xy}(\mathbf{k}) \left[n_{F}(-|\mathbf{d}|)-n_{F}(|\mathbf{d}|) \right]
\end{equation}
where $n_{F}(x)=1/[1+e^{\beta(\mathbf{k}^{2}/2m-\mu+x)}]$, and $Q_{xy}(\mathbf{k})$ is the topological charge given by,  
\begin{equation}
Q_{xy}(\mathbf{k})=\frac{1}{2 \pi} \hat{d}(\mathbf{k})\cdot [\partial_{k_{x}} \hat{d}(\mathbf{k}) \times \partial_{k_{y}} \hat{d}(\mathbf{k})]
\end{equation}
with $\hat{d} \equiv \mathbf{d}/|\mathbf{d}|$. Thus, we obtain, 
\begin{equation} \label{Top_charge}
Q_{xy}(\mathbf{k})=\frac{M \alpha^{2} }{2 \pi (\alpha^{2} \mathbf{k}^{2}+M^{2})^{3/2} }.
\end{equation}

\begin{figure}[H]
	\centering    
	\includegraphics[width=0.47\textwidth]{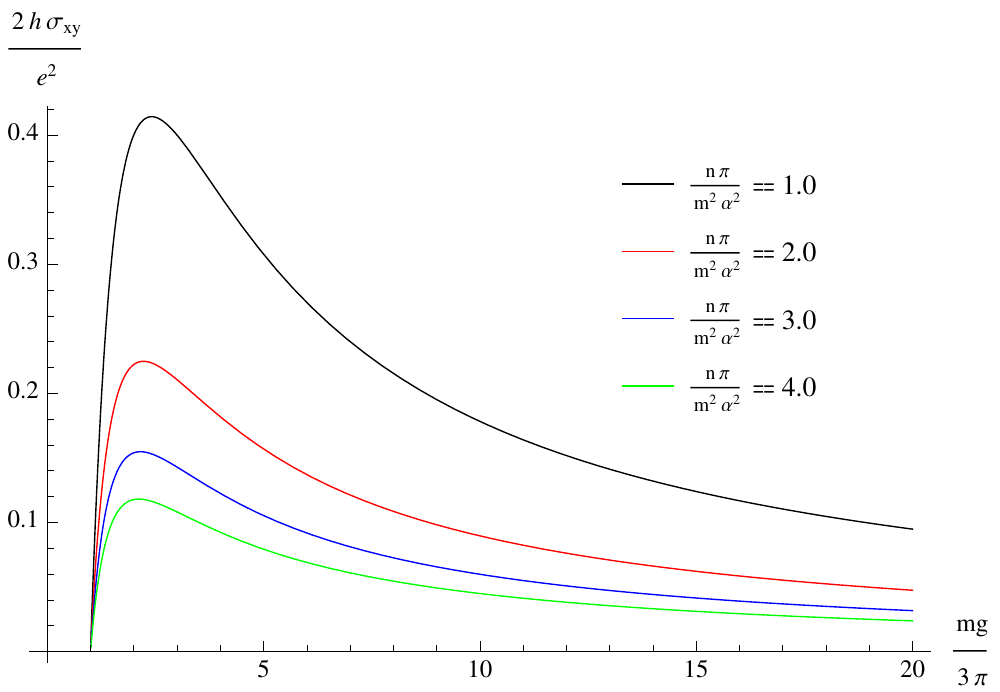}
	\caption{(Color online) Hall conductivity as a function of the coupling constant for $\alpha=1$ and different values of $\frac{\pi n}{m^{2} \alpha^{2}}$ at zero temperature. }
	\label{c_x_M(T=0)}
\end{figure}

\begin{figure}[h]
	\centering    
	\includegraphics[width=0.47\textwidth]{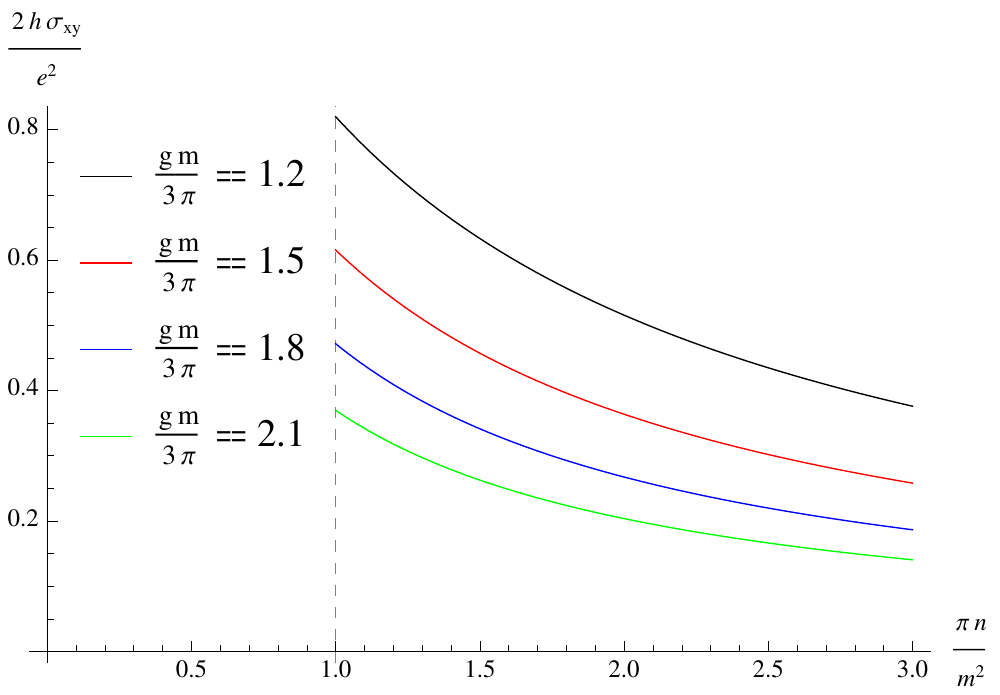}
	\caption{(Color online) Hall conductivity as a function of $\frac{\pi n}{m^{2}}$ for $\alpha=1$ and different values of the coupling constant at zero temperature. }
	\label{c_x_n(T=0)}
\end{figure}

\begin{figure}[H]
	\centering    
	\includegraphics[width=0.47\textwidth]{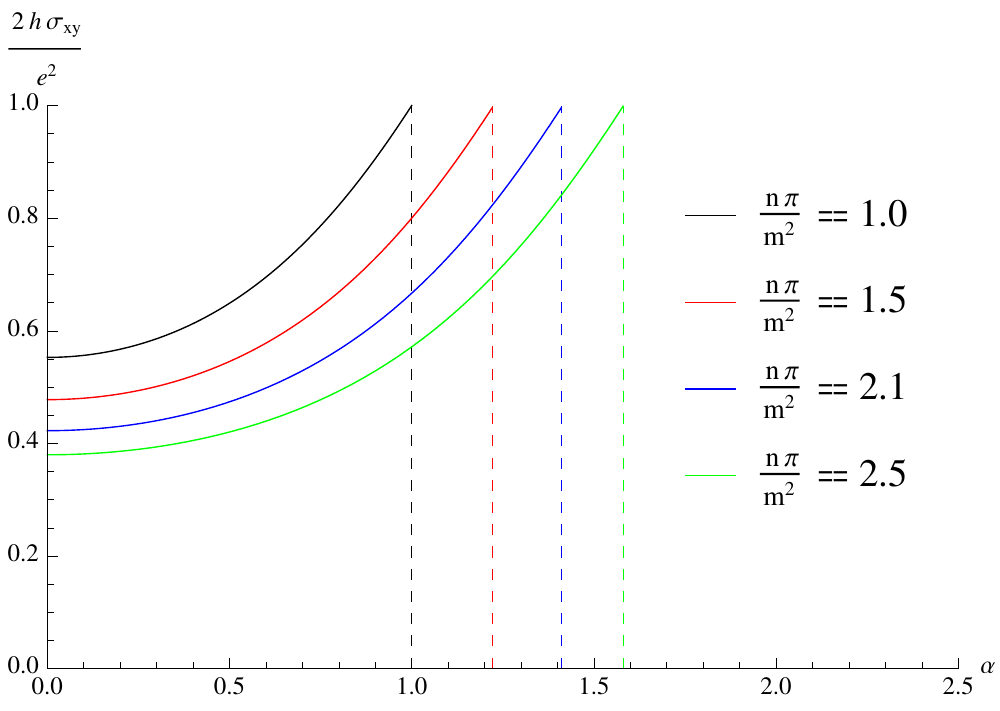}
	\caption{(Color online) Hall conductivity as a function of the spin-orbit coupling $\alpha$ for $\frac{mg}{3 \pi}=1$ and different values of  $\frac{\pi n}{m^{2}}$ at zero temperature. }
	\label{c_x_a(T=0)}
\end{figure}

At zero temperature Eq. (\ref{Hall}), with the topological charge of Eq. (\ref{Top_charge}), can be analytically solved leading to;
\begin{eqnarray} 
\label{Hall(T=0)}
\frac{2h}{e^2}\sigma_{xy}(M,T=0)&=&M\left[ \frac{\Theta \left( 1-|M|/\mu \right)}{\Delta-\alpha^{2}m}-
\frac{1}{\Delta+\alpha^{2}m} \right]   \nonumber\\
&+&{\rm sgn}(M) \Theta \left(|M|/\mu-1 \right).
\end{eqnarray}
where $\Delta=\sqrt{m^2\alpha^4+2m\alpha^2\mu+|M|^2}$. Using Eq. (\ref{M_T=0}) it is easy to see that the Hall conductivity on Eq. (\ref{Hall(T=0)}) is different from zero only in the ferromagnetic insulating phase and actually reduces to, 
\begin{equation} \label{Conductivity}
\frac{2h}{e^2}\sigma_{xy}(M,T=0)=\left( {\rm sgn}(M)-\frac{M}{\Delta+\alpha^{2}m} \right) \Theta (g-g_{c}),
\end{equation}
which is plotted in Fig. \ref{c_x_M(T=0)}, Fig. \ref{c_x_n(T=0)} and Fig. \ref{c_x_a(T=0)}. 

The limit $m\to\infty$ is also straightforward to obtain,
\begin{equation}
\frac{2h}{e^2}\sigma_{xy}(M,T=0)|_{m\to\infty}=\frac{M}{\mu}+\left( \frac{M}{|M|}-\frac{M}{\mu} \right) \theta(|M|-\mu)
\end{equation}
and leads to a broken symmetry metallic solution, see Fig. \ref{c(T=0,m=inf)}. 

\begin{figure}[h]
	\centering    
	\includegraphics[width=0.49\textwidth]{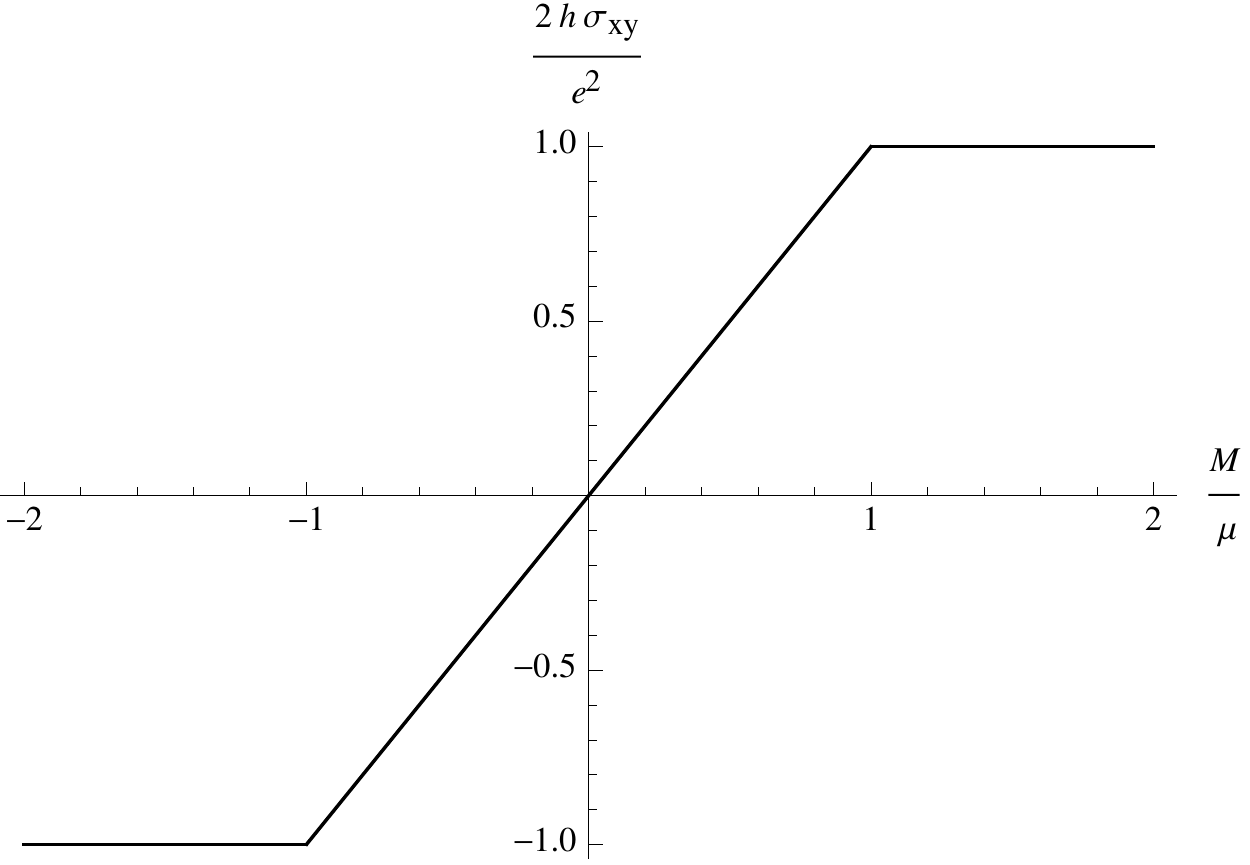}
	\caption{(Color online) Hall conductivity as a function of the ratio between the gap and the chemical potential in the $m \rightarrow \infty$ limit at zero temperature.}
	\label{c(T=0,m=inf)}
\end{figure}

Considering the massive case, the Hall conductivity remains finite only for $g>g_c$ since there is no ferromagnetic metallic phase and although the gap increases with the coupling constant, Eq. (\ref{Conductivity}) ensures the conductivity only grows until a certain maxima and than decreases showing an asymptotic behavior $g \rightarrow \infty \Rightarrow |M| \rightarrow \infty \Rightarrow \frac{2h \sigma_{xy}}{e^2} \rightarrow 0$, as showed in Fig. \ref{c_x_M(T=0)}. For $\alpha=\alpha_{c}=\frac{\pi n}{m^{2}}$ and $g=g_{c}=\frac{3 \pi}{m}$ the magnetization vanishes (consequently the Hall conductivity also), as discussed before and one can see in Fig \ref{c_x_n(T=0)} and Fig \ref{c_x_a(T=0)}. Under the $m \rightarrow \infty$ limit spontaneous TR symmetry breaking is responsible for the appearance of a ferromagnetic metallic solution, in which the Hall conductivity varies linearly with the ratio $|M|/ \mu$, in the insulator phase the hall conductivity remains quantized and assumes the values $\sigma_{xy} =\pm e^{2}/2h$.

Although the effective Hamiltonian in Eq. (\ref{H_mean-field}) is quite general, the particular choice for $\mathbf{d}$ in this paper is related to a continuum model instead of some periodic lattice. For a real lattice insulating model the gap is oppened in such a way that the energy minimum of one band is larger than the maximum of the other band for any momentum in the first Brillioun zone, this ensures that $n_{F}(-|\mathbf{d}|) \rightarrow 1$ and $n_{F}(|\mathbf{d}|) \rightarrow 0$ in the zero-temperature limit and Eq. (\ref{Hall}) leads to a quantized Hall conductivity independent of the details of the lattice. The model presented in this paper may be viewed as some effective theory obtained after an expansion around the gamma point, for example, for a block related to a quantum well Hamiltonian \cite{Quantum_Well}, and features different results from the lattice. First, the unbounded kinetic term $\frac{\mathbf{k}^{2}}{2m}$ ensures that the energy minimum from one band is not always larger than the energy from the other for all different momenta $(\mathbf{k})$. Thus, the band-gap is not fully opened, unlike the case of a lattice insulating model, and the proper zero-temperature limits of the Fermi distributions, necessary for the system to feature a quantization of the Hall conductivity, are not achieved. Second, in the $m \rightarrow \infty$ limit the band is fully gapped and the quantization of the Hall conductivity is achieved in the insulating phase, but leads to half-integer values. This is also expected since the two-band Hamiltonian here is just one block of a whole quantum well Hamiltonian, thus the number of edge states is reduced and consequently the total winding number related to the quantization of the Hall conductivity, as discussed in \cite{Qi(Kubo_formula)}.

On the other way effective theories such as the one presented here may be useful to describe the two-dimensional surface of a three-dimensional topological insulator in which TR symmetry is broken due to the proximity effect to some ferromagnetic material \cite{Nogueira_Hall}, notice that the system in this paper under the $m \rightarrow \infty$ limit is TR invariant only in the paramagnetic phase $(\phi=0)$.  In this case each surface of the topological insulator would give a half-integer contribution to the Hall conductivity, such as the one obtained here.

\section{Conclusion}

We investigated a Hubbard-Rashba model with short-range Coulomb interaction. The Coulomb interaction is decoupled by a saddle-point approximation in the effective action, obtained after a Hubbard-Stratonovich transformation in the spin channel and integration of the fermionic fields. Under this approximation the charge density and the gap equation are analytically solved and the Hall conductivity is computed. The limit of infinite mass, which is a way to consider strong spin-orbit coupling (a common situation for those interested in topological properties) is also considered. 

For finite mass our system always features a first-order metal-insulator phase transition, except in the absence of chemical potential, 
in which case quantum criticality occurs. In the limit of infinite mass the ferromagnetic phase is split into a ferromagnetic insulating phase, which also appears for a finite mass, and a metallic ferromagnetic one (non-existing for finite mass) due to spontaneous TR symmetry breaking. The Hall conductivity is quantized only in the insulating phase in the infinite mass limit. In the metallic region the Hall conductivity remains zero for a finite mass, since the whole metallic phase is paramagnetic, but varies linearly with the ratio $M/\mu$ for $m \to \infty$.

Our results are quite different from the ones obtained with standard models of magnetism such as the Hubbard, Stoner, Heisenberg and Kane-Mele ones, but a first-order metal-insulator transition featuring an important dependence on the density has already been found before in the lattice by Mott \cite{Mott}.
The next step is to investigate the phase transitions of this system using other techniques, such as the renormalization group, and make the same investigation in the case of a cubic Rashba SOI, since recent experiments \cite{Cubic-Rashba} allow for a better investigation of the effects of the cubic Rashba coupling in Ge/SiGe compounds.

\section*{Acknowledgement}

The author thanks Ilya Eremin and Flavio Nogueira from RUB - Ruhr Universit\u{a}t Bochum for very helpful and interesting discussions about this work and general topics on Condensed Matter Physics. The author also acknowledges support from brazilian agency CAPES (CSF 11763/13-2).

\end{document}